\documentclass[12pt]{article}
\usepackage{amsmath,amsthm,amsfonts,amssymb}
\def\R{{\mathbb R}}
\def\Z{{\mathbb Z}}
\def\M{{\mathcal M}}
\def\E{{\mathcal E}}
\def\A{{\mathcal A}}
\def\B{{\mathcal B}}
\def\G{{\mathsf G}}
\def\Diff{{\mathrm{Diff}}}
\def\Hom{{\mathrm{Hom}}}
\def\Conf{{\mathrm{Conf}}}
\def\Vir{{\mathrm{Vir}}}
\def\epsilon{{\varepsilon}}

\newtheorem{theorem}{Theorem}[section]

\makeatletter
\def\section{\@startsection {section}{1}{\z@}{-3.5ex plus -1ex minus
 -.2ex}{2.3ex plus .2ex}{\normalsize\bf}}
\def\@maketitle{\newpage
 \null
 \vskip 2em \begin{center}
 {\large \@title \par} \vskip 1.5em {\normalsize \lineskip .5em
\begin{tabular}[t]{c}\@author
 \end{tabular}\par}
 \vskip 1em {\normalsize \@date} \end{center}
 \par
 \vskip 1.5em}
\makeatother

\title{{\bf Classification of operator algebraic conformal field theories\\
in dimensions one and two}}
\author{{\sc Yasuyuki Kawahigashi}\\
Department of Mathematical Sciences\\
University of Tokyo, Komaba, Tokyo, 153-8914, Japan\\
e-mail: {\tt yasuyuki@ms.u-tokyo.ac.jp}}
\begin{document}
\date{}
\maketitle

\begin{abstract}
We formulate conformal field theory in the setting of algebraic quantum
field theory as Haag-Kastler nets of local observable algebras with
diffeomorphism covariance on the two-dimensional Minkowski space.  We then
obtain a decomposition of a two-dimensional theory into two chiral theories.
We give the first classification result of such chiral theories with
representation theoretic invariants.  That is, we use the central charge as
the first invariant, and if it is less than 1, we obtain a complete
classification.  Our classification list contains a new net which does not
seem to arise from the known constructions such as the coset or 
orbifold constructions.  We also present
a classification of full two-dimensional conformal theories.
These  are joint works with Roberto Longo.
\end{abstract}

\section{Introduction}

Our main results, together with R. Longo, are classification
results for conformal field theories, in the operator algebraic
approach.  We first briefly describe our basic framework for
quantum field theory and its relation to a more conventional approach
based on Wightman axioms using operator-valued distributions.

Our framework is called {\sl algebraic quantum field theory}
or {\sl local quantum physics}, and its standard textbook
is \cite{H} by R. Haag.  We first explain our axiomatic setting on
the 4-dimensional Minkowski space, although we will later
work on lower dimensional spacetime.  Recently, several
attempts have been made on studies on curved spacetime or
even noncommutative spacetime, but we will not deal with
such topics in this review.

In our setting, a physical system is described by a family
of operator algebras $\A(O)$ on a fixed Hilbert space $H$,
where $O$ is a bounded region in the Minkowski space.
As such a region $O$, we consider only 
double cones, which are of the form
$(x + V_+)\cap(y + V_-)$, where $x,y\in \R^4$ and
$$V_\pm = \{ z=(z_0,z_1,z_2,z_3)\in \R^4 \mid
z_0^2 - z_1^2 - z_2^2 - z_3^2 > 0,  \pm z_0 > 0\}.$$
We assume that we have a von Neumann algebra $\A(O)$ acting
on $H$ for each double cone $O$ and the following properties
hold.  (An algebra of bounded linear operators on a Hilbert
space is called a von Neumann algebra if it is closed under
the $*$-operation and weak-operator topology.)
\begin{enumerate}
\item (Isotony) For $O_1\subset O_2$, we have $\A(O_1)\subset \A(O_2)$.
\item (Locality) If $O_1$ and $O_2$ are spacelike separated, then
elements in $\A(O_1)$ and $\A(O_2)$ commute.
\item (Poincar\'e Covariance)
There exists a unitary representation $U$ of
the universal covering of the restricted Poincar\'e group satisfying
$\A(gO)=U_g \A(O) U_g^*$.
\item (Vacuum) We have a unit vector $\Omega\in H$, unique up to
phase, satisfying $U_g \Omega =\Omega$ for all elements $g$ in the
restricted Poincar\'e group and $\bigcup_O \A(O)\Omega$ is dense
in $H$.
\item (Spectrum Condition) If we restrict the representation
$U$ to the translation subgroup, its spectrum is contained in
the closure of $V_+$.
\end{enumerate}

The isotony axioms simply states that we have more observables for
a larger region.  The locality axiom means that if we have
two spacelike separated regions, then we have no interactions
between them even at a speed of light, so the two operators
taken from the two regions mutually commute.  It is also called
the Einstein causality.  Covariance means that
a ``spacetime symmetry'' acts as a symmetry of the family of
operator algebras.  We will later use a higher spacetime symmetry than
restricted Poincar\'e group.  The
vector $\Omega$ is called a vacuum vector and it gives a vacuum
state.  The spectrum condition means stability.

If we denote the set of the elements that are spacelike separated
with all the elements of a region $D$ by
$D^\perp$, then we have $O^{\perp\perp}=O$ for a double cone
$O$.  This is why we use only double cones.  For a general
region $D$, we could define $\A(D)$ as the von Neumann algebra
generated by $\A(O)$ for all double cones $O$ contained in $D$.

Since the set of double cones is directed with respect to 
inclusions, we often say that the family $\A(O)$ is a {\sl
net of von Neumann algebras}.  We also say a {\sl net of
factors}, if each von Neumann algebra $\A(O)$ 
has a trivial center, which is often the case in the lower dimensional
spacetime as below, since
such a von Neumann algebra is called a factor.  
In many such cases, the local algebras 
are all isomorphic, so each local algebra itself does not
contain physical information about the system.
A basic idea is that all
information about a certain physical system is contained in
such a net $\A(O)$.  

 From a mathematical viewpoint, such a net of von Neumann algebras
is simply a family of operator algebras subject to certain set
of axioms, so we can study classification theory of such families
of operator algebras up to an obvious notion of isomorphism.
A useful and important tool for such a study is a representation
theory of a net of von Neumann algebras.

A basic tool to study a net of von Neumann algebras is its
representation theory formulated by Doplicher-Haag-Roberts (DHR)
\cite{DHR}.  Each operator algebra $\A(O)$ acts on a fixed Hilbert
space from the beginning, but we can also consider a representation
of a family of operator algebras on a different Hilbert space where
we do not have a vacuum vector any more.  A basic idea of
Doplicher-Haag-Roberts is that if we assume a nice condition
called the Haag duality and select a nice class of representations
with their criterion, then each such representation is
realized, up to unitary equivalence, as a certain endomorphism
of (the norm closure of) $\bigcup_O \A(O)$.  Such an endomorphism
is often called a DHR endomorphism.  An important feature of
endomorphisms is that they can be composed.  This composition
gives an operation in the set of DHR endomorphisms which plays
a role of a tensor product.  Through this operation (and others),
mathematical structure of DHR endomorphisms  becomes quite 
similar to that of unitary representations of a compact group,
and it gives a $C^*$-tensor category.

We briefly mention a relation of the above approach to a more
conventional one based on the Wightman axioms.
In the setting of the Wightman axioms, one considers a
family of operator-valued distributions $\{\phi_j(x)\}$ on
the Minkowski space.  If we have such a family at the
beginning, then, roughly speaking, we apply smooth
functions supported in $O$ to these distributions, apply
bounded functional calculus to the resulting (unbounded) 
operators, and let $\A(O)$ be the von Neumann algebra generated 
by these bounded operators.  In this way, we should obtain a local
net of von Neumann algebras.  If we start with a local net $\A(O)$
of von Neumann algebras, we should obtain operator valued
distributions $\{\phi_j(x)\}$ through a certain limiting 
procedure in which bounded regions $O$ shrink to one point $x$.
It is believed that the approach
based on the Wightman axioms and  the one based on local nets of von
Neumann algebras are essentially equivalent, and there have been
many works which study under what conditions we obtain one 
from the other, but the exact relations between the two approaches
have not been fully understood yet.

\section{Full and chiral algebraic conformal quantum field theories}

The above general framework in the previous section obviously
works on a Minkowski space of any dimension.  We now specialize
on the 2-dimensional Minkowski space $\M$ and require higher symmetry
than the general Poincar\'e covariance.  This is our approach
to conformal field theory.  Then through a chiral decomposition
of a full algebraic conformal quantum
field theory, we obtain a chiral algebraic conformal quantum
field theory which is now described as a one-dimensional net
of factors.  After such a general description, we will briefly
mention a relation to vertex (operator) algebras, which give
another mathematical approach to chiral conformal field theories.

We now work on a two-dimensional Minkowski space $\M$ where we
use $t$ and $x$ for the time and space coordinates, 
respectively.  We have a von Neumann algebra $\A(O)$ on a
fixed Hilbert space $H$ for each double cone $O$ in this
Minkowski space $M$ as above.  We set $L_\pm=\{t\pm x=0\}$ and
each double cone is a direct product $I_+\times I_-$, where
$I_\pm$ are bounded intervals in $L_\pm$, respectively.
We consider the M\"obius group $PSL(2,\R)$ which
acts on $\R\cup\{\infty\}$ as linear fractional
transformations.  In this way, we obtain a local action
of the universal covering group $\overline{PSL}(2,\R)$
on $\R$.  We impose the following axioms for our
net of von Neumann algebras $\A(O)$ on $H$ and call such a net a
{\sl M\"obius covariant net} of von Neumann algebras.
(See \cite{KL2} for more details.)

\begin{enumerate}
\item (Isotony) For $O_1\subset O_2$, we have $\A(O_1)\subset \A(O_2)$.
\item (Locality) If $O_1$ and $O_2$ are spacelike separated, then
elements in $\A(O_1)$ and $\A(O_2)$ commute.
\item (M\"obius Covariance)
There exists a unitary representation $U$ of
$\overline{PSL}(2,\R)\times
\overline{PSL}(2,\R)$ on $H$ such that for every double
cone $O$, we have $\A(gO)=U_g \A(O) U_g^*$ when $g\in W$ where
$W$ is a connected neighbourhood of the identity in 
$\overline{PSL}(2,\R)\times
\overline{PSL}(2,\R)$ satisfying $gO\subset M$ for
all $g \in W$.
\item (Vacuum) We have a unit vector $\Omega\in H$, unique up to
phase, satisfying $U_g \Omega =\Omega$ for all elements $g$
and $\bigcup_O \A(O)\Omega$ is dense in $H$.
\item (Positive energy)  The one-parameter unitary subgroup
of $U$ corresponding to the time translation has a positive
generator.
\end{enumerate}

We now further strengthen the axiom of M\"obius covariance as follows.
Let $\G$ be the quotient of $\overline{PSL}(2,\R)\times
\overline{PSL}(2,\R)$ modulo the relation
$(r_{2\pi}, r_{-2\pi})=({\rm id}, {\rm id})$.  Then it turns
out that our representation $U$ as above gives a representation
of this group $\G$, due to the spacelike locality.  We then find
that our net $\A(O)$ extends to a local $\G$-covariant net on
the Einstein cylinder $\E=\R\times S^1$, which is the cover of the
2-torus obtained by lifting the time coordinate from $S^1$ to $\R$.
We also have several consequences from the above set of axioms.
See \cite[Proposition 2.2]{KL2}, for example.

Let $\Diff(\R)$ be the group of the orientation preserving
diffeomorphisms which are smooth at infinity.  Then this group
naturally acts on $\E$ as a diffeomorphic action.
Let $\Conf(\E)$ be the group of global, orientation
preserving conformal diffeomorphisms of $\E$.  This group is
generated by $\Diff(\R)\times \Diff(\R)$ and $\G$.  If
a M\"obius covariant net $\A$ further satisifies the following
axiom, we say that the net $\A$ is a {\sl local conformal net}.
This is the class we study.

(Diffeomorphism covariance) The unitary representation $U$ of $\G$
extends to a projective unitary representation of $\Conf(\E)$
such that the extended net on $\E$ is covariant.  Furthermore,
we have $U_g X U_g^*=X$ for $g\in \Diff(\R)\times\Diff(\R)$,
if $X\in \A(O)$ and $g$ acts on $O$ as identity.

This gives our framework for conformal quantum field theory.  
We study a net $\A(O)$ as a family of von Neumann algebras
satisfying the above set of axioms.  Such a family here is
also called a {\sl full algebraic conformal quantum field theory}.  
The DHR theory works in this setting perfectly.

Suppose we have a local conformal net $\A$ as above.
Then for each bounded interval $I\subset L_+$, we set
$\A_+(I)=\bigcap_J \A(I\times J)$.  In this way, we have
a family of von Neumann algebras $\A_+$ parameterized by
bounded intervals $I$.  We regard these von Neumann algebras
as subalgebras of $B(H_+)$, where $H_+$ is the closure of
$\bigcup_I \A_+(I)\Omega$.  This family extends to a
family $\A_+(I)$, where $I$ is any open, nondense, nonempty,
and connected set of $S^1=\R\cup \{\infty\}$.  (Such $I$ is
simply called an interval in $S^1$.)
This family $\A_+(I)$ satisfies the following conditions.
We may take these as axioms for such a family.  (See \cite{GL}
for more details.)

\begin{enumerate}
\item (Isotony) For $I_1\subset I_2$,
we have $\A_+(I_1)\subset \A_+(I_2)$.
\item (Locality) If $I_1$ and $I_2$ are disjoint, then
elements in $\A_+(I_1)$ and $\A_+(I_2)$ commute.
\item (Diffeomorphism Covariance)
There exists a projective unitary representation $U$ of
$\Diff(S^1)$ on $H_+$ such that for every interval $I$,
we have $\A_+(gI)=U_g \A_+(I) U_g^*$.
Furthermore, we have $U_g X U_g^*=X$ for $g\in \Diff(S^1)$,
if $X\in \A(I)$ and $g$ acts on $I$ as identity.
\item (Vacuum) We have a unit vector $\Omega\in H_+$, unique up to
phase, satisfying $U_g \Omega =\Omega$ for all elements $g$
in the M\"obius group $PSL(2,\R)$
and $\bigcup_I \A_+(I)\Omega$ is dense in $H_+$.
\item (Positive energy)  The one-parameter unitary subgroup
of $U$ corresponding to the rotation on $S^1$ has a positive
generator.
\end{enumerate}

A family $\A_+$ satisfying the above set of axioms is
called a {\sl chiral algebraic conformal quantum
field theory}.  We can similarly
define $\A_-$.  We then have an embedding $\A_+(I)\otimes \A_-(J)
\subset \A(I\times J)$.  The DHR theory works fine for a chiral
algebraic conformal quantum
field theory.  Now, for two DHR endomorphisms 
$\rho, \sigma$, we have a unitary equivalence for
$\rho\sigma$ and $\sigma\rho$, but we have a canonical
unitary $\epsilon(\rho,\sigma)$ implementing this unitary
equivalence and this family $\epsilon$ of unitaries contains
non-trivial information.  This family $\epsilon$ is called
a {\sl braiding}.  See \cite{FRS}, for example, for details.

Our classification method for full algebraic conformal quantum
field theories
$\A(O)$ consists of two steps.  In the first step, we classify
the chiral algebraic conformal quantum
field theories $\A_\pm$.  In the second
step, we classify the embedding $\A_+(I)\otimes \A_-(J)
\subset \A(I\times J)$, which is a non-trivial subfactor, usually.

\section{Complete rationality and classification}

Our basic idea for classification is that if we have a certain
nice condition, generally called ``amenability'', a simple
set of invariants related to representation theory should give
a complete classification.  We have given a general idea along
this line in \cite{K}, so here we only briefly explain the
condition called {\sl complete rationality}, which was introduced
in \cite{KLM} and plays a role of amenability in classification
theory. 

Here we state complete rationality for a chiral algebraic conformal
quantum field theory $\A(I)$, $I\subset S^1$.  We also have a version
for a full algebraic conformal quantum
field theory and we refer the reader to
\cite{KL2} for the definition in such a setting.
Consider a chiral algebraic conformal quantum
field theory $\A$.  Split $S^1$
into $2n$ intervals, and label them $I_1, I_2,\dots,I_{2n}$
in the counterclockwise order.  Let $\mu_n$ be the
Jones index of the subfactor
$\A(I_1)\vee\A(I_3)\vee\cdots\vee\A(I_{2n-1})\subset
(\A(I_2)\vee\A(I_4)\vee\cdots\vee\A(I_{2n}))'$.  (Note that
we have this inclusion because of locality.)  This number is
independent of the way to split the circle.  We remark
that we automatically have $\mu_1=1$, which is called the
{\sl Haag duality}.  Complete rationality consists of the 
following three conditions.

\begin{enumerate}
\item (Strong additivity) Remove one point from an interval $I$ and
label the resulting two intervals as $I_1, I_2$.  Then
we have $\A(I)=\A(I_1)\vee\A(I_2)$.
\item (Split property) Consider two intervals $I_1, I_2$ with
$\bar I_1 \cap \bar I_2=\emptyset$.  Then the von Neumann algebra
$\A(I_1)\vee\A(I_2)$ is naturally isomorphic to
$\A(I_1)\otimes\A(I_2)$.
\item (Finiteness of the $\mu$-index) We have $\mu_2<\infty$.
\end{enumerate}

The main results in \cite{KLM} give the following two conditions
under complete rationality.

\begin{enumerate}
\item We have only finitely many equivalence classes of irreducible
DHR endomorphisms of the net $\A$.
\item The braiding naturally gives a unitary representations of
$SL(2,\Z)$ whose dimension is the number of the equivalence classes
in (1).
\end{enumerate}

This shows that the category of the DHR endomorphisms of the net
$\A$ gives a modular tensor category, which plays an important
role in theory of quantum invariants of 3-manifolds as in \cite{T}.

We next explain the first numerical invariant of a local
conformal net, a central charge, of $\A$.  Let $\A$ be
a local conformal net.  (Here we do not need complete
rationality.)  Then we have a projective unitary
representation of $\Diff(S^1)$.
Recall that the Virasoro algebra is the infinite dimensional Lie
algebra generated  by elements $\{L_n \mid n\in\Z\}$ and $c$ with
relations
$$[L_m,L_n]=(m-n) L_{m+n} + \frac{c}{12}(m^3-m)\delta_{m,-n},$$
and $[L_n,c]=0$.  This is unique, non-trivial one-dimensional 
central extension of the Lie algebra of $\Diff(S^1)$.
We now obtain a representation of the Virasoro algebra and
then the central element $c$ is mapped to a scalar.  This value
is the {\sl central charge} of the net $\A$ and is also
denoted by $c$.  It has been shown by
Friedan-Qiu-Shenker
\cite{FQS} that this central charge value is in
$$\{1-6/m(m+1)\mid m=2,3,4,\dots\}\cup[1,\infty)$$ and
the values $\{1-6/m(m+1)\mid m=2,3,4,\dots\}$ have been 
realized by Goddard-Kent-Olive
\cite{GKO}.  (The values in $[1,\infty)$ are easier to
realize.)   Jones has proved in his theory of index for
subfactors \cite{J} that the index value is in the
set $\{4\cos^2\pi/m\mid m=3,4,5\dots\}\cup[4,\infty]$
and all the values in this set can be realized.
It is obvious that we have a formal similarity between the
two cases.  A relation between the Jones theory of subfactors
and algebraic quantum field theory was found in \cite{L1}.
Our classification results give further deeper relations between
the two.  (See \cite{EK} for a general theory of subfactors and
related topics.)

In classification theory of subfactors, Ocneanu \cite{O1}
has found a paragroup, which gives a combinatorial
invariant for a subfactor through its representation
theory.  If the Jones index
value is less than 4, the subfactor is of finite depth
automatically, and this finite depth case is a special
case of the amenable case which Popa's classification
theorem \cite{P1} covers.
We have shown in \cite{KL1} that if the central charge value
is less than 1, then the net of factors is automatically
completely rational. 

Wassermann's construction of the $SU(n)_k$ nets based on
loop group representations gives the first examples of chiral
algebraic conformal quantum field theories and they are
completely rational.
The finiteness of the $\mu$-index for these nets was
proved by Xu \cite{X2}.  Xu also studied the coset and orbifold
constructions in the setting of chiral algebraic conformal quantum
field theory in \cite{X3,X4}.  They give completely rational
nets by \cite{X4,L3}.

We briefly note that we have some formal similarity between
our complete rationality and a condition in 
theory of vertex operator algebras, which is another
mathematical approach to a chiral conformal field theory.
They have a condition
called {\sl $C_2$-finiteness} introduced by Zhu \cite{Z}, which
is formally analogous to the above finiteness
of the $\mu$-index.  See \cite{FBZ}
for more details on vertex operator algebras.
On a vertex operator algebra $V$, we have binary operations
$a_{(n)} b$, $a, b\in V$, parameterized by integers $n$.  The
finiteness of the codimension $\dim V/V_{(-2)} V$ is called
the $C_2$-finiteness condition.  A vertex operator algebra $V$
is said to be {\sl rational} if every $V$-module is completely
reducible, and this condition implies that $V$ has only finitely
many inequivalent simple modules.  Zhu has proved that if
we have the $C_2$-finiteness condition, then the modular
group $SL(2,\Z)$ acts on the space of characters of all the
mutually inequivalent simple $V$-modules.
This finiteness of the codimension and the above finiteness of the
$\mu$-index have some formal common similarity as follows.
\begin{enumerate}
\item The $\mu$-index is also a certain multiplicative codimension.
\item Both the codimension $\dim V/V_{(-k)} V$ and the index
$\mu_k$ can be defined for any positive integer $k$.
\item The above codimension and the index are trivial for $k=1$.
\item If the above codimension and the index are finite for 
$k=2$, then they are also finite for all positive integers $k$,
and we obtain a unitary actions of the modular group $SL(2,\Z)$
on certain natural finite dimensional spaces.
\end{enumerate}
However, the action of  $SL(2,\Z)$ in the setting of vertex
operator algebras is on the space of characters while its
action in the setting of representation categories of a 
net of factors is on the intertwiner spaces, and we do not have
any direct relation between the two situations.  It would be
very interesting to clarify this formal analogy.

\section{$\alpha$-induction, modular invariants, and classification}

Here we explain our classification method for completely rational
nets on $S^1$ with central charge less than 1.

Suppose we have a chiral algebraic conformal quantum
field theory $\A(I)$, $I\subset S^1$,
with central charge $c<1$.  Then the projective representation $U$ of
the diffeomorphism group gives a subnet as follows.
For an interval $I\subset S^1$, we define $\B(I)$ be the von
Neumann algebra generated by $U_g$, where $g$ is a diffeomorphism
which acts trivially outside of $I$.  It is easy to see that
this $\B(I)$ is a subalgebra of $\A(I)$ by the Haag duality and
$\B$ gives a subnet in the sense of \cite{L3}.  (Note that the
vacuum vector is not cyclic for $\B$.)  We use a notation
$\Vir_c(I)$ for this subnet and call it a Virasoro subnet
with central charge $c$.  This net is among the coset constructions
due to Xu \cite{X3}, which relies on A. Wassermann's construction of
$SU(2)_k$-nets \cite{W}.  (See \cite{C} for more on the
Virasoro nets.)  We have shown in \cite{KL1} that
the subfactor $\Vir_c(I)\subset \A(I)$ has a trivial
relative commutant and finite index, which is a quite nontrivial
fact.

For a net of subfactors $\Vir_c\subset\A$, we have a machinery
of $\alpha$-induction, which is analogous to a machinery
of induction and restriction for representations of groups.
For a DHR endomorphism $\lambda$ of the smaller net $\Vir_c$,
we obtain an endomorphism $\alpha^\pm_\lambda$ of the larger
net $\A$.  This is ``almost'' a DHR endomorphism, but not
completely.  This operation is regarded as an extension of
an endomorphism and depends on a choice $\pm$ of the
braiding on the system of DHR endomorphisms of the smaller 
net.  This method was defined by Longo-Rehren \cite{LR}
and many interesting properties and examples were studied
by Xu \cite{X1} and B\"ockenhauer-Evans \cite{BE}.  Ocneanu
\cite{O2} had a graphical method based on a quite
different motivation, and it was unified with the theory
of $\alpha$-induction by us in \cite{BEK1,BEK2,BEK3}.
One of the main results in \cite{BEK1} is that if
we define a matrix $Z$ by $Z_{\lambda\mu}=\dim\Hom(\alpha^+_\lambda,
\alpha^-_\mu)$ for irreducible DHR endomorphisms $\lambda,\mu$
of the smaller net, then this matrix is in the commutant of
the unitary representation of $SL(2,\Z)$ arising from the
braiding on the system of DHR endomorphisms of the smaller net.
Obviously, each entry of $Z$ is a nonnegative integer and
we have $Z_{00}=1$ for the vacuum representation denoted by
$0$.  Such a matrix $Z$ is called a {\sl modular invariant}
(of the unitary representation of $SL(2,\Z)$).

It is easy to see that for a given unitary representation
of $SL(2,\Z)$, we have only finitely many modular invariants,
and this finite number is often quite small in concrete examples.
For the Virasoro net $\Vir_c$, the corresponding modular
invariants have been completely classified by Cappelli-Itzykson-Zuber
\cite{CIZ} and they are labeled with pairs of $A$-$D$-$E$ Dynkin
diagrams with difference of the Coxeter numbers being 1.  Also
it is fairly easy to see in our current context that we have
only so-called type I modular invariants in the classification
of \cite{CIZ} where we have only $A_n$, $D_{2n}$, $E_6$, $E_8$
diagrams.  In this way, starting with a chiral algebraic conformal 
quantum field theory $\A$ with central charge less than 1, we
obtain a type I modular invariant matrix $Z$ in the classification
list of \cite{CIZ} labeled with pairs of the
$A_n$-$D_{2n}$-$E_{6,8}$ Dynkin diagrams
with difference of the Coxeter numbers being 1.  Our main
result in \cite{KL1} with R. Longo is that this correspondence
gives a complete classification of  chiral algebraic conformal 
quantum field theories.  Note that we have no reason, a priori,
to believe or expect that this correspondence from a conformal
field theory to a matrix in a certain list
is injective or surjective, but we have proved both 
injectivity and surjectivity of this  correspondence.

Our classification list is as follows.

\begin{enumerate}
\item Virasoro nets with central charge $c=1-6/m(m+1)$.
\item Their simple current extensions of index $2$.
\item The exceptional net labeled with $(E_6, A_{12})$.
\item The exceptional net labeled with $(E_8, A_{30})$.
\item The exceptional net labeled with $(A_{10}, E_6)$.
\item The exceptional net labeled with $(A_{28}, E_8)$.
\end{enumerate}

The first two of the above exceptional ones  are realized as
the coset constructions for
$SU(2)_{11}\subset SO(5)_1\otimes SU(2)_1$ and
$SU(2)_{29}\subset (G_2)_1\otimes SU(2)_1$.  They were first
considered by B\"ockenhauer-Evans
\cite[II, Subsection 5.2]{BE} as possible candidates
realizing the corresponding modular invariants in the
Cappelli-Itzykson-Zuber list, but they were unable to show
that these coset constructions indeed produce the desired modular
invariants.  With our complete classification, it is easy to
identify these cosets with the above two in our list.

Recently, K\"oster \cite{Ko}
identified the third exceptional net in the above list ,
$(A_{10}, E_6)$, with the two cosets $SU(9)_2\subset (E_8)_2$ and
$(E_8)_3\subset (E_8)_2\otimes (E_8)_1$, assuming that the local
conformal nets $(E_8)_k$ have the expected WZW-fusion
rules.  The last one, $(A_{28}, E_8)$, does not seem to be
a coset nor an orbifold, and it appears to be a genuine new example.

Carpi \cite{C} and Xu \cite{X5} recently obtained
certain classification results of chiral algebraic conformal
quantum field theories with central charge equal to 1,
independently.

\section{Classification of 2-dimensional theories and 2-cohomology}

We now explain how to obtain a classification of 
full algebraic conformal quantum field theories with central
charge less than 1, using the results in the previous section.
This is our joint work with R. Longo \cite{KL2}.
As we mentioned before, our strategy is to study a subfactor
$\A_+(I)\otimes \A_-(J) \subset \A(I\times J)$, where $\A(I\times J)$
is a given full algebraic conformal quantum field theory with
central charge 1.  By the classification list in the previous
section, we have a complete information on the chiral ones
$\A_\pm$.

We now assume the so-called parity symmetry condition for a full
algebraic conformal quantum field theory $\A$, which in particular
implies that $\A_+$ and $\A_-$ are isomorphic and they contain
the same $\Vir_c$.
Then the dual canonical endomorphism for the subfactor
$\Vir_c(I)\otimes \Vir_c(J) \subset \A(I\times J)$ gives a decomposition
$\bigoplus_{\lambda,\mu} Z_{\lambda\mu} \lambda\otimes\mu$,
where $\lambda,\mu$ are irreducible DHR endomorphisms of $\Vir_c$.
In a more general setting, the following was conjectured by Rehren and
proved by M\"uger \cite{Mu}.

\begin{theorem}
Under the above conditions, the following are equivalent.
\begin{enumerate}
\item The net $\A$ has only the trivial representation theory.
\item The $\mu$-index of the net $\A$ is 1.
\item The matrix $Z$ above is a modular invariant.
\end{enumerate}
\end{theorem}

In this way, we obtain a modular invariant $Z$ for $\Vir_c$ in
the classification list of Cappelli-Itzykson-Zuber \cite{CIZ}
from a full algebraic conformal quantum field theory $\A$
with parity symmetry and trivial representation theory.
Then we can prove as in \cite{KL2} that a full algebraic
conformal quantum field theory $\A$ with parity symmetry 
has only the trivial representation theory if and only if
it is maximal with respect to extensions.  Thus we obtain
a modular invariant in \cite{CIZ} from full algebraic conformal
quantum field theory $\A$ with parity symmetry and maximality.
Our main result in \cite{KL2} with R. Longo shows that this
gives a bijective correspondence.  Note that the modular
invariants in \cite{CIZ} are labeled with pairs of $A$-$D$-$E$ Dynkin
diagrams with difference of the Coxeter numbers being 1 as before,
but we now do not have a restriction to so-called type I modular
invariants, so the Dynkin diagrams $D_{2n+1}$ and $E_7$ do appear.

Surjectivity of this correspondence is not
difficult by Rehren's result \cite{R}, together with
our previous analysis in \cite{BEK2,KL1}.

To prove injectivity of this correspondence, we need to study
the subfactor $\A_+(I)\otimes \A_-(J) \subset \A(I\times J)$,
where we have a natural identification of $\A_+(I)$ and $\A_-(J)$
and the dual canonical endomorphism decomposes in the form
of $\bigoplus \lambda\otimes\lambda$, where $\lambda$ is
in a system of irreducible endomorphisms.  In subfactor theory,
such a subfactor was first studied by Ocneanu \cite{O1} under
the name of the asymptotic inclusion and it plays a role of 
the quantum double construction.  (See \cite{EK} and also
\cite{BEK3} for more details.)
Popa \cite{P2} gave a quite general construction
and named it a symmetric enveloping algebra.  Here we use
a formulation of Longo-Rehren \cite{LR} where they gave a 
specific system of $Q$-system in the sense of Longo \cite{L2}.

Based on \cite{LR}, we studied this type of subfactors in \cite{KL2}
and found that the $Q$-system of  a general subfactor of this type 
has a ``twist'' arising from a 2-cocycle of the tensor category
of the endomorphisms.  This notion of a 2-cocycle for a tensor
category is a generalization of a 2-cocycle for a finite group,
so it does not vanish in general, but we have proved in 
\cite{KL2} it always vanishes for the tensor categories of
the representation categories of chiral algebraic conformal
quantum field theories with central charge less than 1.
This 2-cohomology vanishing gives the desired injectivity
of the above correspondence from a full algebraic conformal
quantum field theory to the modular invariants in \cite{CIZ}.

With extra combinatorial work along the same line, we can
drop the triviality condition of the representation theory
and  classify full algebraic conformal
quantum field theories with parity symmetry and central
charge less than 1 completely as in \cite{KL2}.

\section*{Acknowledgments}

We gratefully acknowledge the support of
GNAMPA-INDAM and MIUR (Italy) and
Grants-in-Aid for Scientific Research, JSPS (Japan).
We thank R. Longo, A. Matsuo, and  C. Schweigert for
useful correspondences on the related materials.

\end{document}